\documentclass[cup7b]{cupbookJR}

\usepackage{makeidx}         
\usepackage{graphicx}        
\usepackage[bottom]{footmisc}

\usepackage{dsfont}
\usepackage{color, url}
\usepackage[latin2]{inputenc}
\usepackage{amsmath,mathrsfs}
\usepackage{amssymb,amsbsy}
\usepackage{epsfig} 

\newtheorem{theorem}{Theorem}[section]

\newcommand{\rem}[1]{}

\rem{

}

\newcommand{\non}{\nonumber}

\newcommand\etc{etc.\ }

\newcommand{\bfi}{\bfseries\itshape}

\DeclareMathAlphabet{\mathbi}{OML}{cmm}{b}{it} 
\newcommand{\bx}{\mathbi{x}}

\newcommand{\bel}{\begin{equation}\label}
\newcommand{\ee}{\end{equation}}
\newcommand{\ben}{\begin{enumerate}}
\newcommand{\een}{\end{enumerate}}
\newcommand{\bde}{\begin{description}}
\newcommand{\ede}{\end{description}}
\newcommand{\bit}{\begin{itemize}}
\newcommand{\eit}{\end{itemize}}
\newcommand{\bc}{\begin{center}}
\newcommand{\ec}{\end{center}}
\newcommand{\bB}{\mbox{\boldmath$B$}}
\newcommand{\bH}{\mbox{\boldmath$H$}}

\newcommand{\bdB}{\mbox{\boldmath$\mathcal{B}$}}
\newcommand{\bdD}{\mbox{\boldmath$\mathcal{D}$}}

\newcommand{\bu}{\mathbi{u}}

\newcommand{\bU}{\mbox{\boldmath$\mathcal{U}$}}

\newcommand{\bom}{\mbox{\boldmath$\omega$}}

\newcommand{\bsD}{\boldsymbol{D}}
\newcommand{\bsE}{\boldsymbol{E}}
\newcommand{\bsH}{\boldsymbol{H}}

\newcommand{\bcapom}{\mbox{\boldmath$\Omega$}}

\newcommand{\bk}{\mbox{\boldmath$\hat{k}$}}

\newcommand{\beq}{\begin{eqnarray}} 
\newcommand{\eeq}{\end{eqnarray}}

\newcommand{\bS}{\mathbi{S}}

\begin{document}
\author[]{J. D. Gibbon and D. D. Holm\\
Department of Mathematics, Imperial College London,\\
London, SW7 2AZ. UK. \\
\texttt{j.d.gibbon@ic.ac.uk~and~d.holm@ic.ac.uk}
}
\chapter[Stretching \& folding diagnostics for fluids]{\textbf{\Large Stretching \& folding 
diagnostics in solutions of the three-dimensional Euler \& Navier-Stokes\\equations}}

\bigskip

\begin{abstract}
Two possible diagnostics of stretching and folding (S\&F) in fluid flows are discussed, based on the 
dynamics of the gradient of potential vorticity ($q = \bom\cdot\nabla\theta$) associated with 
solutions of the three-dimensional Euler and Navier-Stokes equations.  The vector $\bdB = \nabla q
\times \nabla\theta$ satisfies the same type of stretching and folding equation as that for the 
vorticity field $\bom $ in the incompressible Euler equations (Gibbon \& Holm, 2010). The quantity 
$\theta$ may be chosen as the potential temperature for the stratified, rotating Euler/Navier-Stokes 
equations, or it may play the role of a seeded passive scalar for the Euler equations alone. The 
first discussion of these S\&F-flow diagnostics concerns a numerical test for Euler 
codes and also includes a connection with the two-dimensional surface quasi-geostrophic equations. 
The second S\&F-flow diagnostic concerns the evolution of the Lamb vector $\bsD = \bom\times\bu$, 
which is the nonlinearity for Euler's equations apart from the pressure. The curl of the Lamb 
vector ($\boldsymbol{\varpi} := \mbox{curl}\,\bsD$) turns out to possess similar stretching and 
folding properties to that of the $\bdB$-vector. 
\end{abstract}
\par\vspace{3mm}
\bc
{\small PACS numbers: 47.10.A-,\,47.15.ki}
\ec


\section{\large Introduction}\label{intro}

This paper considers two variants of the stretching and folding properties of gradients of solutions 
of the three-dimensional Euler and Navier-Stokes equations, following recent work of Gibbon \& Holm 
(2010). Fine-scale structures, diagnosed in either inviscid or viscous turbulence and MHD  by the 
presence of large gradients, are created in the tortuous stretching and folding processes that arise 
from the vortex stretching term in the Euler and Navier-Stokes equations.  These fine-scale structures 
are not wholly understood, as they lie at the heart of unsolved regularity issues that have challenged 
mathematicians for more than a generation. 
\par\smallskip
In what follows the advected scalar field $\theta$ will comprise either\,: 
(i) the potential temperature for the stratified, rotating Euler/Navier-Stokes equations\,; 
or (ii) a passive scalar for the Euler equations alone. The main theme revolves around the 
role of the vector $\bdB = \nabla q\times \nabla \theta$ where the potential vorticity $q = 
\bom\cdot\nabla\theta$ is conserved on fluid particle paths in either case. The basis of 
the result, already discussed by Kurgansky \& Tatarskaya (1987) and Kurgansky \& Pisnichenko 
(2000), is that in the incompressible Euler case the vector $\bdB$ satisfies the same equation 
as that for the vorticity, thus suggesting intriguing stretching and folding properties for 
the gradient of the projection of $\bom$ on the normal to level surfaces of $\theta$. This 
result, summarized in \S\ref{incomNS}, also has interesting consequences for the Navier-Stokes 
equations (Gibbon \& Holm, 2010).  
\par\smallskip
The first of the variants on this theme in \S\ref{eultest} concerns a scheme for testing 
numerical Euler codes which have been designed to address the issue of whether the equations 
develop a finite time singularity. Thus it is apposite to devote the introduction to this 
section \S\ref{eulsing} to listing some of the Euler literature in this area. A closely 
associated problem forms the subject of \S\ref{2DQG} in which a connection is established 
with the two-dimensional surface quasi-geostrophic equations (2D-QG) studied by Constantin, 
Majda \& Tabak (1994). In this $\nabla^{\perp}\theta$ in two-dimensions obeys the same vortex 
stretching equation as that of $\bom$ for three-dimensional Euler.  It is shown that the 
2D-QG equations are embedded as a special case in the equation for $\bdB$.  
\par\smallskip
The second main variant revolves around the stretching and folding properties of the Lamb 
vector $\bsD = \bom\times\bu$ for the incompressible Euler equations. The Lamb vector 
comprises the nonlinearity of the Euler equations aside from the pressure, so its evolution 
is of importance. In \S\ref{incomEul} it is shown that its curl 
\bel{varpidef}
\boldsymbol{\varpi} := \mbox{curl}\,\bsD = \mbox{curl}\,
(\bom\times\bu)
\ee
also satisfies the same type of stretching equation as that for $\bdB$, while its divergence 
($\mbox{div}\bsD$) obeys a continuity equation. 
In the compressible case this may have interesting consequences 
for the study of jet-noise although this is beyond the present scope of this paper. Section \ref{Helden} 
deals with the evolution of the gradient of helicity density $\lambda = \bom\cdot\bu$ which also appears to 
possess similar stretching and folding properties.
\par\medskip
Let us begin with the notation for the incompressible Euler equations, which are 
expressed as
\bel{3Deul1a}
\frac{D\bu}{Dt} = -\nabla p\,,
\quad {\rm div}\,\bu = 0\,,
\ee
or as 
\bel{3Deul1b}
\partial_{t}\bu - \bu\times\bom = -\nabla \big(p + \tfrac12 u^{2}\big)\,.
\ee
The chosen domain is a three-dimensional periodic box $\mathcal{V} = [0,L]^{3}$. $\bu$ is 
the velocity field of the fluid and the material derivative is defined by
\bel{matdef}
\frac{D~}{Dt} = \partial_{t} + \bu\cdot\nabla\,.
\ee
The vorticity field $\bom = \mbox{curl}\,\bu$ satisfies
\bel{3Deul2}
\partial_{t}\bom - \mbox{curl}(\bu\times\bom) = 0\,.
\ee
This formula can also be written in the familiar vortex stretching format
\bel{3Deul3}
\frac{D\bom}{Dt} = \bom\cdot\nabla\bu \equiv \textbf{S}\,\bom\,,
\ee 
where $\rm{S}_{ij} = \tfrac12\left(u_{i,j}+u_{j,i}\right)$ is the rate of strain matrix.  Equations 
(\ref{3Deul2}) and (\ref{3Deul3}) are equivalent evolution equations for $\bom$.  Euler data roughens 
very quickly, a fact which is mainly due to the \textbf{stretching and folding processes} 
caused by the rapid alignment or anti-alignment of $\bom$ with positive and negative 
eigenvectors of $\textbf{S}$. 

The main aim of this paper is to show that these stretching 
and folding processes are shared by several other variables in the Euler and Navier-Stokes 
equations.
\par\smallskip
While the existence of some very weak solutions has been proved (Shnirelman, 1997; 
Brenier, 1999; Majda \& Bertozzi, 2001; De Lellis \& Sz\'ekelyhidi, 2007, 2008; 
Brenier, De Lellis \& Sz\'ekelyhidi, 2009), nevertheless Leray-type weak solutions 
are unknown. However, if we are to progress in our understanding of the properties 
of solutions of the Euler equations, our lack of knowledge forces us to make some 
assumptions about the existence of solutions. The fundamental result on existence of 
solutions of the three-dimensional Euler equations is the theorem due to Beale, Kato 
\& Majda (1984) which is assumed to hold\,:
\begin{theorem}\label{BKMthm} (Beale, Kato \& Majda, 1984) There exists a global solution 
$\bu \in C([0,\,\infty];H^{s})\cap C^{1}([0,\,\infty];H^{s-1})$ of the Euler equations for 
$s\geq 3$ if and only if, for every $t > 0$, 
\bel{BKMequn1}
\int_{0}^{t}\|\bom(\cdot,\,\tau)\|_{\infty}\,d\tau < \infty \,.
\ee 
\end{theorem}
\textbf{Remarks.} (i) The value of this result is that computationally only the 
quantity $\|\bom\|_{\infty}$ needs to be monitored. If this is finite everywhere in 
the domain of flow at a time $t$ then the solutions are regular at that time.
\par\smallskip\noindent
(ii) It does not predict a singularity in $\|\bom\|_{\infty}$ but it restricts those that may 
potentially occur of the type $\|\bom\|_{\infty}\sim (t_{s} - t)^{-p}$ to the range $p\geq 1$. 
When $p < 1$ the theorem is violated. 
\par\smallskip\noindent
(iii) Kozono and Taniuchi (2000) have proved a version of this theorem in the BMO-norm (bounded 
mean oscillations) which is slightly weaker than the $L^{\infty}$-norm.
\par\smallskip\noindent
(iv) Further analytical approaches have centred around conditional estimates on the magnitude 
and direction of vorticity that include the direction of vorticity. These are extensions of the 
Beale-Kato-Majda theorem\,; the most significant papers are those by Constantin, Fefferman \& 
Majda (1996), Deng, Hou \& Yu (2005,\,2006) and Chae (2003,\,2004,\,2005,\,2007).

\section{The $\bdB$ equation for the stratified Euler \& Navier-Stokes equations}\label{incomNS}

\subsection{The stratified, rotating Euler equations}\label{sreul}

Let us consider the three-dimensional incompressible Euler equations for an 
incompressible, stratified, rotating flow ($\bcapom = \bk\,\Omega$) in terms 
of the velocity field $\bu(\bx,\,t)$ and the potential temperature $\theta$
\bel{ens1}
\frac{D\bu}{Dt} + 2\,(\bcapom\times\bu) + a_{0}\bk\,\theta = -\nabla p\,,
\ee
where $a_{0}$ is a dimensionless constant and where $\theta(\bx,\,t)$ evolves 
passively according to 
\bel{ens2}
\frac{D\theta}{Dt} = 0\,.
\ee
How $\theta(\bx,\,t)$ and other variables might accumulate into large local 
concentrations is of interest\footnote{The BKM theorem expressed in the last 
section is valid when $\theta$ is no more than a passive scalar driven by an 
Euler flow as in (\ref{3Deul2}). For stratified Euler (\ref{ens1}) together 
with (\ref{ens2}), however, 
it is necessary to assume that $\int_{0}^{t}\left(\|\bom\|_{\infty} + 
\|\nabla\theta\|_{\infty}\right)\,d\tau$ is finite.}. 
To pursue this, consider the vorticity 
$\bom = \mbox{curl}\,\bu$ and define $\bom_{rot} = \bom + 2\bcapom$, which satisfies
\bel{ens3}
\frac{D\,\bom_{rot}}{Dt} = \bom_{rot}\cdot\nabla\bu - \nabla^{\perp}\theta
\qquad\qquad\nabla^{\perp} = \left(\partial_{y},\,-\partial_{x},\,0\right)
\,.\ee
The potential vorticity defined by (Hoskins, McIntyre \& Robertson, 1985) 
\bel{ens4}
q = \bom_{rot}\cdot\nabla\theta,
\ee
satisfies Ertel's theorem  (Ertel, 1942; Truesdell \& Toupin, 1960; 
Ohkitani, 1993; Kuznetsov \& Zakharov, 1997) because 
\beq\label{ens5}
\frac{Dq}{Dt} &=& \left(\frac{D\,\bom_{rot}}{Dt} - \bom_{rot}\cdot\nabla\bu \right)\cdot\nabla \theta
+ \bom_{rot}\cdot\nabla\left(\frac{D\theta}{Dt}\right)\non\\
&=& -\nabla^{\perp}\theta\cdot\nabla \theta = 0\,.
\eeq
\par\vspace{-4cm}
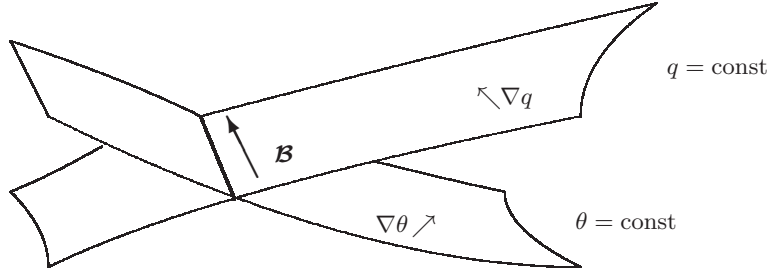
\begin{figure}[htb]\hspace{-3.5in}
\begin{minipage}[htb]{1cm}\setlength{\unitlength}{1cm}
\begin{picture}(11,11)(0,0)
%
\qbezier[400](1,5)(5,3)(8,3)
\qbezier[400](1,5)(.75,5.5)(.5,6)
\qbezier[400](.5,6)(2,5.5)(3,5)
\qbezier[400](3,5)(6,5.8)(9,6.5)
\qbezier[400](8,3)(7,3.5)(7,4)
\qbezier[400](7,4)(6.5,4.1)(5.3,4.4)
\qbezier[400](1,3)(3,4)(8,5)
\qbezier[400](1,3)(1,3.5)(.5,4)
\qbezier[400](.5,4)(1,4.2)(1.7,4.6)
\qbezier[200](9,6.5)(8,5.8)(8,5)
\thicklines
\qbezier[400](3.45,3.9)(3.2,4.5)(3,5)
\put(3.75,4.2){\vector(-1,2){.4}}
\put(4.1,4.4){\makebox(0,0)[b]{$\bdB$}}
\put(8.6,3.5){\makebox(0,0)[b]{$\theta = \mbox{const}$}}
\put(9.8,5.5){\makebox(0,0)[b]{$q = \mbox{const}$}}
\put(5.7,3.4){\makebox(0,0)[b]{$\nabla\theta\!\nearrow$}}
\put(7,5.1){\makebox(0,0)[b]{$\nwarrow\!\!\nabla q$}}
\end{picture}
%
\par\vspace{-25mm}\noindent
\end{minipage}
\caption{\label{Fig-qt-int} 
The vector $\bdB$ points along an intersection of level sets of the two Lagrangian flow 
constants $(q,\,\theta)$}.
\end{figure}
This establishes two quantities $q$ and $\theta$ that are each conserved along flow lines, 
and whose level sets intersect as in Figure \ref{Fig-qt-int}. Then with $\bdB$ defined as 
\bel{ens6a}
\bdB = \nabla q \times\nabla\theta
\ee
Kurgansky \& Tartskaya (1987), Kurgansky \& Pisnichenko (2000) have observed that $\bdB$ 
satisfies\footnote{Note that this is a Clebsch representation of the divergence-free vector 
$\bdB$, not a decomposition of the vorticity $\bom$. See Ohkitani (2008) for a recent study of 
the latter and Holm \& Kupershmidt (1983) for a review of the Clebsch approach.
Moreover, the helicity of $\bdB$ given by $\int_\Omega \bdB\cdot\,{\rm curl}^{-1} \bdB\,dV$ 
is necessarily zero for homogeneous or periodic boundary conditions.}
\bel{ens6b}
\partial_{t}\bdB = \mbox{curl}\,(\bu\times\bdB)\,.
\ee
Of course this may be written equivalently in the familiar vortex stretching format (\ref{3Deul3})
\bel{ens6c}
\frac{D\bdB}{Dt} = \bdB\cdot\nabla\bu
\ee
thereby highlighting the fact that alignment of $\bdB$ with eigenvectors of $\nabla\bu$ is 
critical to the stretching process. In Figure \ref{Fig-qt-int} the vector $\bdB$ is tangent to the 
curve defined by the intersection of $q = \mbox{const}$ and $\theta = \mbox{const}$. Thus,   
$\bdB$ plays the same role as that for $\bom$ and for the magnetic $\bB$-field 
in MHD (Moffatt, 1978; Palmer, 1988). Hence, all the stretching and folding properties 
associated with vorticity or magnetic field-lines also apply to $\bdB$ even though 
$\bdB$ contains various projections of $\bom$, $\nabla\bom$, $\nabla\theta$ and $\nabla
\nabla\theta$. Moreover, for any surface $S(\bu)$ moving with the flow $\bu$, one finds
\bel{ens7}
\frac{d}{dt}\int_{S(\tiny\bU)} \bdB \cdot dS = 0\,.
\ee

\subsection{The stratified Navier-Stokes equations}\label{NSE}

Now let us turn to the Navier-Stokes equations coupled to the $\theta$-field. (In what 
follows the rotation will be ignored.) These equations are
\bel{ens8a}
\frac{D\bu}{Dt} + a_{0}\theta\,\bk = Re^{-1}\Delta\bu - \nabla p\,,
\ee
\bel{ens8b}
\frac{D\theta}{Dt} = \big(\sigma Re\big)^{-1}\Delta\theta\,.
\ee
Here, the potential vorticity $q = \bom\cdot\nabla\theta$ is no longer a material constant 
but, instead, evolves according to
\beq\label{ens9}
\frac{Dq}{Dt} &=& \left(\frac{D\bom}{Dt} - \bom\cdot\nabla\bu \right)\cdot\nabla \theta
+ \bom\cdot\nabla\left(\frac{D\theta}{Dt}\right)\non\\
&=& \big(Re^{-1}\Delta\bom - \nabla^{\perp}\theta \big)\cdot\nabla \theta
+ \bom\cdot\nabla\left[(\sigma Re)^{-1}\Delta\theta\right]
\non\\
&=& \mbox{\bf\sf div}\big\{Re^{-1}\Delta\bu\times\nabla \theta 
+ (\sigma Re)^{-1}\bom\Delta\theta \big\}\,.
\eeq
The material advection property is destroyed but the introduction of a \textit{transport velocity 
field} $\bU_{q}$ transforms (\ref{ens9}) into a continuity equation
\bel{ens10}
\partial_t q +  \mbox{div}\,(q\,\bU_{q}) = 0\,,
\ee
thus making $q$ a PV density, and where $\bU_{q}$ is defined through the relation
\bel{ens11}
q\big(\bU_{q} - \bu\big) = - Re^{-1}\big(\Delta\bu\times\nabla \theta  + 
\sigma^{-1}\bom\Delta\theta\big)\,.
\ee
The introduction of the transport velocity field $\bU_{q}$ is originally due to Haynes \& 
McIntyre (1987). Note that $\mbox{div}\,\bU_{q} \ne 0$ although $\mbox{div}\,\bU_{q} = O\left( 
Re\right)^{-1}$. Consistent with numerical studies on reconnection (Herring, Kerr \& 
Rotunno 1994), this scaling with Reynolds number $Re$ may suggest that in the early or 
intermediate stages of a flow this divergence may be small.
What about the evolution of the variable $\theta$? It is easily seen that 
\beq\label{ens12}
\partial_{t}\theta + \bU_{q}\cdot\nabla\theta &=& \partial_{t}\theta + \bu\cdot\nabla\theta - 
Re^{-1}q^{-1}\left\{\Delta\bu\times\nabla \theta + \sigma^{-1}\bom\Delta\theta\right\}
\cdot\nabla\theta\non\\
&=& \partial_{t}\theta + \bu\cdot\nabla\theta - \big(\sigma Re\big)^{-1}\Delta\theta 
= \textbf{0}\,.
\eeq
The formal result for the stratified Navier-Stokes equation is\,:
\begin{theorem}\label{Bthm}
The scalar quantities $q$ and $\theta$ satisfy
\bel{ens13}
\partial_{t} q + \mbox{div}\,\big(q\,\bU_{q}\big) = 0\,,\qquad \partial_{t}\theta + 
\bU_{q}\cdot\nabla\theta = 0\,,
\ee
and $\bdB = \nabla q\times\nabla \theta$ satisfies the stretching and folding relation
\bel{ens14}
\partial_{t}\bdB - \mbox{curl}\,(\bU_{q}\times\bdB) = \bdD_{q}\,,
\ee
where the divergence-less vector $\bdD_{q}$ is given by 
\bel{ens15}
\bdD_{q} = - \nabla(q\,\mbox{div}\,\bU_{q})\times\nabla\theta\,,
\ee
and the transport velocity $\bU_{q}$ is defined as in (\ref{ens11}). Moreover, for any 
surface $\bS(\bU_{q})$ moving with the flow $\bU_{q}$
\bel{ens16}
\frac{d}{dt}\int_{\bS(\tiny\bU_{q})} \bdB \cdot d\bS = \int_{\bS(\tiny\bU_{q})}\bdD_{q}\cdot d\bS\,.
\ee
\end{theorem}
This is the natural way of expressing problems in the vortex stretching format using the transport 
velocity $\bU_{q}$.

\section{The Euler singularity problem}\label{eulsing}

Out of large-scale computations of solutions of the three dimensional Euler equations has 
emerged the natural question of whether a singularity develops in a finite time (Majda \& 
Bertozzi, 2001; Bardos \& Titi, 2007; Constantin, 2008; Gibbon, 2008). An extensive 
literature has arisen on this question but no conclusion has yet been agreed\,: see 
Bardos \& Benachour (1977); Morf, Orszag \& Frisch (1980); Chorin (1982), Brachet, Meiron, 
Orszag, Nickel, Morf \& Frisch (1983); Siggia (1984); Kida (1985); Ashurst \& Meiron (1987); 
Pumir \& Kerr, (1987); Pumir \& Siggia (1990); Grauer \& Sideris (1991); Bell \& Marcus (1992); 
Brachet, Meneguzzi, Vincent, Politano \& Sulem (1992); Kerr (1993,\,2005a,\,2005b); Boratav \& 
Pelz (1994,\,1995); Pelz (1997,\,2001); Pelz \& Gulak (1997); Grauer, Marliani \& Germaschewski 
(1998); Cichowlas \& Brachet (2005); Gulak \& Pelz (2005); Pelz \& Ohkitani (2005); Pauls, Matsumoto, 
Frisch \& Bec (2006). Regarding more recent work, the fine-scale computations by Hou and Li (2006,\,2007) 
that see only super-exponential growth in $\bom$ contradict both the older computations of Kerr 
(1993,\,2005a) together with 
newer results by Bustamante and Kerr (2007), in which a finite time singularity has been 
observed. Similar but not identical anti-parallel vortex tube initial conditions have been 
used in the two bodies of results which largely coincide until a late stage. Two further 
recent contributions are those of Orlandi and Carnevale (2007) who used initial conditions 
in the form of Lamb dipoles to observe singular behaviour, as did Grafke, Homann, 
Dreher \& Grauer (2007). 

\subsection{A numerical test for Euler computations}\label{eultest}

Let us take the Euler equations for the velocity field $\bu$ in their standard form as in 
(\ref{3Deul1a}) without the rotation or buoyancy used in \S\ref{incomNS}. The new feature of 
the proposed test is to introduce a passive tracer concentration $\theta(\bx,\,t)$ satisfying 
\bel{theta1}
\frac{D\theta}{Dt} = 0\,,
\ee
and  whose 
initial data are under the investigator's control.
Introducing $\theta$ allows us, as before, to use $q=\bom\cdot\nabla\theta$ which 
still obeys 
\bel{q1}
\frac{Dq}{Dt} = 0\,,
\ee
and which therefore allows the use of the same definition $\bdB = \nabla q \times\nabla \theta$.
This is endowed with initial conditions inherited from those for $\bu$ and $\theta$. $\bdB$ must 
satisfy 
\bel{B1}
\partial_{\,t}\,\bdB = \mbox{curl}\,{(\bu\times\bdB)}\qquad\mbox{or}\qquad
\frac{D\bdB}{Dt} = \bdB\cdot\nabla\bu\,,
\ee
which mimics the vorticity stretching equation (\ref{3Deul2}). The critical point about the 
use of the vector field $\bdB$ is that it has embedded information on $\bom$, $\nabla\bom$, 
$\nabla\theta$ and $\nabla\nabla\theta$. It evolves in the same way as $\bom$ and so is subjected 
to similar stretching and folding processes. It can, however, be evaluated 
at any particular time $t$ in several distinct ways\,: it can be evaluated from the result of the 
evolution in (\ref{B1}) at time $t$, or it can be computed from its definition using $\bu$ and 
$\theta$ evolved up to and evaluated at time $t$. The degree to which these distinct evaluations 
agree or disagree provides a quantitative gauge of the accuracy of the numerical computation. It 
is not clear that there is a natural scale for the inevitable discrepancies produced in any particular 
computation. However, this procedure produces a precise diagnostic quantity that, given identical 
initial data, can be directly compared side-by-side for different numerical computations to evaluate 
their {\em relative} accuracy. The suggested test is\,:
\ben\itemsep 0mm
\item Choose initial data for $\bu$ and $\theta$, thereby fixing initial data for $q$ and $\bdB$. 

\item Evolve $\bu$ and simultaneously solve $D\theta/Dt = 0$, $Dq/Dt = 0$ and $D\bdB/Dt = \bdB\cdot\nabla\bu$.

\item Test the resolution at any time $t>0$ by constructing $q_{1}(\cdot\,,\,t) =
\bom(\cdot\,,\,t)\cdot\nabla\theta(\cdot\,,\,t)$ and then\,:

\ben
\item compare the solution for $\bdB(\cdot,\,t)$ obtained from solving the stretching equation 
$D\bdB/Dt = \bdB\cdot\nabla\bu$ with
\bel{B1def}
\bdB_{1}(\cdot\,,\,t) = \nabla q_{1}(\cdot\,,\,t) \times \nabla \theta(\cdot\,,\,t)
\ee

\item Furthermore, compare this with 
\bel{B2def}
\bdB_{2}(\cdot\,,\,t) = \nabla q\,(\cdot\,,\,t) \times 
\nabla\theta(\cdot\,,\,t)
\ee
where $q\,(\cdot\,,\,t)$ is the evolved solution of $Dq/Dt=0$.

\een
\item For fixed initial data for $\bu$ this procedure may be implemented for a variety of ``markers''
$\theta_{n}(\cdot\,,\,t)$ evolving from distinct initial data $\theta_{n}(\cdot\,,\,0)$ to diagnose
the numerical accuracy in different regions of the flow. 
\een
Because $\bdB$ contains $\nabla\bom$, comparing the different computations of $\bdB,~\bdB_{1}$
and $\bdB_{2}$ tests the accuracy of the computation of some of the small scale structures in 
the flow. Given the generally acknowledged difficulties in accurately computing the evolution 
of passive scalars such as in (\ref{theta1}) or (\ref{q1}), how initial data are chosen may be 
critical to the calculation. In particular, the appearance of null points in the vorticity 
field may create significant obstacles -- see Ohkitani (2008).

\subsection{Connection with the two-dimensional quasi-geostrophic equations}\label{2DQG}

The open nature of whether the three dimensional Euler equations develops a singularity naturally 
leads to the idea of studying other simpler problems that mimic this behaviour. The foremost 
example is the case of the two dimensional surface quasi-geostrophic (2D-QG) equations. The 
strong fronts observed in numerical computations though the existence of a vortex stretching term 
have led Constantin, Majda \& Tabak (1994) to suggest that these might model singularity development 
in the three dimensional Euler equations.  It turns out that the 2D-QG equations are embedded 
in the equation for $\bdB$ in the following way.
\par\smallskip
Let $q = z = \mbox{const}$ and $\theta = \mbox{const}$ be material surfaces. Then $\bdB$ 
becomes
\bel{qg1}
\bdB = \nabla z \times \nabla\theta = \bk\times\nabla\theta = -\nabla^{\perp}\theta\,.
\ee
So far the velocity field has been left $\bu$ free but if this is then chosen such that 
in $\mathbb{R}^{2}$ 
\bel{qg2}
\bu = -\nabla^{\perp}\psi\qquad\mbox{with}\qquad\theta = -(-\Delta)^{1/2}\psi
\ee
then the equations for $\bdB$ with this $\bu$ satisfies 
\bel{qg3}
\frac{D\bdB}{Dt} = \bdB\cdot\nabla\bu\,.
\ee
These are the 2D-QG equations discussed by Constantin, Majda \&  Tabak (1994) who linked the 
formation of a singularity to the a presence of hyperbolic saddle (for the level sets). 
C\'ordoba (1998) then showed the absence of a 
singularity in the case of a \textit{simple} hyperbolic saddle. We refer the reader 
to for related work connected to the formation of sharp fronts and their evolution\,: 
Ohkitani \& Yamada (1997) Constantin, Nie \& Schorghofer (1998), C\'ordoba (1998), 
C\'ordoba, Fefferman \& Rodrigo (2004) \& Rodrigo (2004).

\section{Transport equations for the curl and divergence of the Lamb vector}\label{lamb}

For both the incompressible and compressible Euler equations, apart from 
the non-local effects of the pressure, the nonlinearity is the \textit{Lamb vector} 
\bel{lambdef}
\bsD = \bom\times\bu\,.
\ee
This is the cross product of vorticity and velocity and is therefore an indicator of regions 
of a flow field where vorticity is nonzero. It shares its critical points with 
velocity and vorticity, but possesses additional critical points where the flow is locally 
Beltrami, as in the helical motion of a swirling jet. 
The evolution of $\bsD$ is of interest\,: its divergence, for example, plays a 
role in the production of jet-noise in compressible flows, wherever the mean of 
${\rm div}\,\bsD\neq 0$. For recent discussions of the utility of the this vector 
as a diagnostic in fluid dynamics or as an important source of jet noise, see 
Rousseaux, Seifer, Steinberg \& Wiebel (2007), Hamman, Klewick \& Kirby (2008) and Cabana, 
Fortun\'e \& Jordan (2008), respectively. In the incompressible case, existence of solutions 
is assured provided the Beale, Kato, Majda (1984) condition is fulfilled. 

\subsection{The evolution of $\bsD$ for the incompressible Euler equations}\label{DincomEul} 

In the case of the incompressible Euler equations, we distinguish the Lamb vector $\bsD = 
\bom\times\bu$ from the {\bfi Bernoulli vector}
\bel{lvzero}
\bsE = \bsD + \nabla(p + \tfrac12 u^{2})\,.
\ee
The incompressible Euler fluid equations (\ref{3Deul1b}) are connected to $\bsD$ and $\bsE$ by 
\bel{lv1a}
\partial_{t}\bu  =  - \,\bsD - \nabla(p + \tfrac12 u^{2}) = -\, \bsE\,,
\ee
which vanishes for steady flows to create Lamb surfaces, reviewed, e.g., in Sposito (1997).
The Bernoulli vector $\bsE$ is distinguished from the Lamb vector $\bsD$ by its divergence, in that
\bel{divE}
{\rm div}\,\bsE = 0\,,
\quad\hbox{while in general}\quad
{\rm div}\,\bsD \ne 0\,,
\ee
although they both share the same curl, i.e., ${\rm curl}\,\bsE={\rm curl}\,\bsD=\boldsymbol{\varpi}$. 
We now choose to rewrite the vorticity equation (\ref{3Deul2}) as 
\bel{lv1b}
\partial_{t}\bom + {\rm curl}\,\bsE = 0\,,
\ee
and thereby remove any gauge freedom in the ${\rm curl}^{-1}$ operation. That is, we choose 
the \emph{Bernoulli gauge}, in which ${\rm curl}^{-1}\boldsymbol{\varpi}=\bsE$.
\par\smallskip
The aim of this section is to show that the curl of the Lamb 
vector $\boldsymbol{\varpi}=\mbox{curl}\,\bsE$ plays a role similar to that of $\bdB$ in 
Theorem \ref{Bthm} and its divergence $\mbox{div}\,\bsD$ obeys a conservation equation that 
introduces an augmented transport velocity field, in the same spirit as in Haynes \& 
McIntyre (1987). 
\par\smallskip
The Euler fluid equations imply the following equation for the evolution of the Lamb vector,
\begin{eqnarray}\label{ell-dot}
\partial_{t}\bsD - \bu\times \boldsymbol{\varpi} &=& \bsE\times\bom\,,
\end{eqnarray}
and so
\begin{theorem}\label{varpithm}
$\boldsymbol{\varpi} = {\rm curl}\,\bsE$ satisfies the stretching equation
\begin{eqnarray}\label{curl-ell-dot}
\partial_{t}\boldsymbol{\varpi} - {\rm curl}\,(\bu\times \boldsymbol{\varpi}) &=& \bdD_{lam}\,,
\end{eqnarray}
where $\bdD_{lam}$ is defined by \[\bdD_{lam} =  {\rm curl}\,(\bsE\times \bom),\] 
and ${\rm div}\,\bsD$ satisfies the conservation equation
\bel{cons-div-ell-1}
\partial_{t}\left({\rm div}\,\bsD\right) + {\rm div}\big[\,\bU\,({\rm div}\bsD)\,\big] = 0\,,
\ee
where the transport velocity $\bU$ is defined by
\bel{jetUdef}
\left(\bU - \bu\right) {\rm div}\,\bsD = 
\bu \times (2\textbf{S}\cdot\bom) + \bom\times\nabla (p + \tfrac12 u^{2})\,,
\ee
and $\textbf{S}$ is the strain-rate tensor. 
\end{theorem}
\par\smallskip\noindent
\textbf{Remarks\,:} This theorem is similar in spirit to Theorem \ref{Bthm} 
for the evolution of $\bdB$ and the continuity equation for $q$, with $\boldsymbol{\varpi}$ 
and $\mbox{div}\,\bsD$ playing these roles respectively. Four further observations about 
these equations follow that all hinge of the property that ${\rm div}\,\bsE = 0$.
\par\medskip\noindent
(1) Another expression for the divergence of the Lamb vector is 
\bel{jet1}
{\rm div}\bsD = -\,\Delta(p + \tfrac12 u^{2})\,. 
\ee
Therefore, equation (\ref{cons-div-ell-1}) is conceivably  
interesting as an evolution equation for the Bernoulli function 
$\left(p + \tfrac12 u^{2}\right)$. 
In compressible turbulence, such as in the exhaust of a jet airplane, the jet noise is largely 
due to correlations that produce a mean ${\rm div}\bsD\ne0$, as discussed in Cabana, Fortun\'e 
\& Jordan (2008). That is, the divergence of the Lamb vector is the leading source of turbulent 
jet noise, so the conservation equation (\ref{cons-div-ell-1}) for its evolution in the 
incompressible case may be of interest. 
The quantity in square brackets in (\ref{cons-div-ell-1}) is the \emph{current 
density} for the transport of the \emph{hydrodynamic charge density}, ${\rm div}\bsD = -\,
\Delta\left(p + \tfrac12 u^{2}\right)$. 
\par\medskip\noindent
(2) Because ${\rm div}\,\bsE = 0$, the Helmholtz equation  (\ref{lv1b}) and 
the curl of (\ref{curl-ell-dot}), rewritten as
\begin{eqnarray}\label{curl-D-dot}
\boldsymbol{\varpi}_t  - {\rm curl}\,(\bu\times \boldsymbol{\varpi}) &=& 
{\rm curl}\,(\bsE\times\bom)\,,
\end{eqnarray}
imply the following two-component system of commutator equations
\begin{eqnarray}\label{curl-ell-dot-com}
\boldsymbol{\varpi}_t + [\bu,\,\boldsymbol{\varpi}] 
&=& [\bom,\,{\rm curl}^{-1}\boldsymbol{\varpi}],\\
\partial_{t}\bom + [\bu,\, \bom] &=& 0\,,\quad\mbox{where}\quad
\bu = {\rm curl}^{-1}\bom
\label{omega-eqn}
\end{eqnarray}
and we have $\bsE = {\rm curl}^{-1}\boldsymbol{\varpi}$ in the Bernoulli gauge. 
The bracket $[\,\cdot\,,\,\cdot\,]$ in these equations denotes commutator of 
divergence-free vector fields. For example,
\bel{lv2}
[\bom,\,\bsE] := \boldsymbol{\omega}\cdot\nabla \bsE -
\bsE\cdot\nabla \boldsymbol{\omega}.
\ee
\par\medskip\noindent
(3) Given that ${\rm div}\,\bsE=0$, one may compute the evolution of $\bsE$-\emph{helicity}, 
defined as 
\bel{lv3}
\Lambda_E := \int \bsE\cdot d\mathbf{x} \wedge d(\bsE\cdot d\mathbf{x})
= \int \bsE \cdot \,{\rm curl}\,\bsE\, dV\,.
\ee
Equations (\ref{curl-ell-dot}) and (\ref{ell-dot}) imply that the helicity of 
$\bsE$ is not constant. Instead, $\Lambda_E$ evolves as
\bel{lv4}
\frac{d}{dt}\int \bsE \cdot \,{\rm curl}\,\bsE\, dV
= -\,2 \int \boldsymbol{\omega} \cdot\left(\bsE \times {\rm curl}\,\bsE\right)\,dV\,,
\ee
after integrating by parts to remove gradient terms and applying homogeneous boundary conditions. 
\par\medskip\noindent
(4) Finally, we remark that the steps taken to distinguish between the fields $\bsD$ 
and $\bsE$ in selecting the Bernoulli gauge \etc are all reminiscent of a formal  
Eulerian analogy with Maxwell's equations for electromagnetism. All that remains in 
completing that well-known formal analogy is to identify $\bom$ with the magnetic field 
$\bB$ and require the curl of the magnetic induction $\bH$ to vanish, i.e., ${\rm curl}\,\bH=0$. 
Then one may interpret  $\partial_{t}\bsD$ in equation (\ref{ell-dot}) as the displacement 
current, the right hand side as the current density, \etc. This is slightly different from 
the variant of that formal analogy discussed previously in Marmanis (1998) and pursued in 
fluid experiments by Rousseaux, Seifer, Steinberg \& Wiebel (2007). For completeness, we 
list the comparison in Table \ref{Table-Max}. 
\begin{figure}[htb]
\begin{tabular}{|l||l|l||l|l|}
\hline
{Maxwell's equations} &{Marmanis (1998)}&{Present paper}\\
\hline\hline
Magnetic Field, $\bB$ &  $\bom ={\rm curl}\,\bu$   &  $\bom ={\rm curl}\,\bu$ 
\\ \hline
Magnetic Induction, $\bsH$ &  Absent    &  $\nabla \chi$  
\\ \hline
Electric Field, $\bsE$ &  $\bom\times\mathbf{u}$ & 
$\bom\times\bu + \nabla(p+\tfrac12 u^2)$ &       
\\ \hline
Displacement vector, $\bsD$ &   Absent   &  $\bom\times\bu$  &       
\\ \hline
Charge density, $q_E$ &  ${\rm div}\,(\bom\times\bu)$  & ${\rm div}\,(\bom\times\bu)$ \\
\hline\hline
\end{tabular}
\caption{\label{Table-Max} Compared with Marmanis (1998), the current Maxwell-hydrodynamics analogy 
distinguishes between $(\bsE,\bB)$ and $(\bsD,\bsH)$. }
\end{figure}
\par\smallskip\noindent
We shall refrain, however, from following the formal analogy between hydrodynamics and 
Maxwell equations here, and finish by interpreting the transport theorems for the Lamb 
vector's divergence (\ref{cons-div-ell-1}) and its curl (\ref{curl-D-dot}) in the standard 
fluid context. 
\par\medskip
In the fluid context, we interpret the commutator equations (\ref{curl-ell-dot-com}) and 
(\ref{omega-eqn}) as evolution equations for the Lagrangian fluxes $\boldsymbol{\varpi}
\cdot d\bS$ and $\bom\cdot d\bS$ as they are swept along by the fluid velocity $\bu$. 
Namely,
\begin{eqnarray}\label{varpi-flux-eq}
\frac{d}{dt} (\boldsymbol{\varpi}\cdot d\bS) 
&=& {\rm curl}\,\big(\bsE \times\bom \big)\cdot d\bS\\
\frac{d}{dt} (\bom\cdot d\bS) &=& 0\,,\label{Helmholtz}
\end{eqnarray}
both along $d\bx(t)/t = \bu(\bx(t),t)$. 
As always, the Helmholtz equation (\ref{Helmholtz}) states that 
the flux of vorticity is frozen into the flow. However, the flux 
of the cross product of vorticity and the Bernoulli vector drives 
the flux of the the Lamb vector's curl, which in turn drives 
the evolution of the vorticity. In a nonlinear feedback response, the 
Lamb vector's curl is driven itself in equation (\ref{varpi-flux-eq}) 
by the flux of the vector product of the vorticity with the Bernoulli 
vector, which contains \emph{both} the nonlinearity and the pressure 
gradient. This process is akin to a magnetic dynamo, with the vorticity 
playing the role of the magnetic field. 
\par\smallskip
One may also express this process as a pair of linked circulation 
theorems, namely, 
\begin{eqnarray} \label{E-circ-eq}
\frac{d}{dt} \oint_{c(\bu)}\bsE\cdot d\bx
&=&
\oint_{c(\bu)}\big(\bsE \times\bom \big)\cdot d\bx\\
\frac{d}{dt} \oint_{c(\bu)} \bu \cdot d\bx &=& 0\,,
\end{eqnarray}
where $c(\bu)$ is a closed material loop moving with the fluid velocity $\bu$. 
Thus, the circulation of the Lamb vector (or the Bernoulli vector) is driven 
by the circulation of the cross product of the Bernoulli vector with the 
vorticity.
\par\smallskip
In addition, the Lamb vector's divergence (${\rm div}\, \bsD$) satisfies 
the continuity equation (\ref{jetUdef}) with its  augmented transport 
velocity that depends on the vorticity, the strain-rate tensor and the 
pressure gradient. 

\subsection{Helicity density}\label{Helden}

Having looked at the dynamics of the Lamb vector $\bsD = \bom\times\bu$ let us consider the 
helicity density of the Euler equations which is the scalar product
\bel{lamdef}
\lambda = \bom\cdot\bu\,.
\ee
Straightforward differentiation gives its dynamical equation,
\bel{lam1}
\frac{D\lambda}{Dt} =  -\bom\cdot\nabla\left(p -\tfrac12 u^{2}\right)
\ee
which may be rewritten equivalently as
\bel{lam2}
\partial_{t}\lambda + 
\mbox{div}\,\left\{\lambda\bu + \bom\left(p - \tfrac12 u^{2}\right)\right\} = 0\,.
\ee
As in \S\ref{incomNS}, this leads to the definition of a transport velocity field $\bU_{\lambda}$ 
\bel{lam3}
\lambda(\bU_{\lambda} - \bu) =  \bom\left(p - \tfrac12 u^{2}\right)
\ee
and the continuity equation
\bel{lam4}
\partial_{t}\lambda + \mbox{div}(\lambda\,\bU_{\lambda}) = 0\,.
\ee
Thus, the vector quantity
\bel{lam5}
\bdB_{\lambda} = \nabla \lambda\times\nabla\theta
\ee
satisfies the stretching and folding result of Theorem \ref{Bthm} of \S\ref{incomNS} 
\bel{lam6}
\partial_{t}\bdB_{\lambda} - \mbox{curl}\,(\bu\times\bdB_{\lambda}) = \bdD_{\lambda}\,,
\ee
with vector $\bdD_{\lambda}$ defined as 
\bel{lam7}
\bdD_{\lambda} = - \nabla\left(\lambda\,\mbox{div}\,\bU_{\lambda}\right)\times\nabla\theta\,.
\ee
The vector $\bdD_{\lambda}$ measures the ``permeability'' or rate of slippage of level 
sets of helicity density through level sets of the passive scalar field, $\theta$.

\section{Conclusion}\label{conclusion}

The stretching and folding processes that produce small-scale structures in either fluid 
turbulence or MHD have generally been associated with the alignment or anti-alignment 
of either the vorticity $\bom$ or the magnetic field $\bB$ with eigenvectors of the 
velocity gradient matrix $\nabla\bu$. The observations and calculations in this paper 
have shown that these stretching and folding processes occur quite widely\,: on the one 
hand they have been shown to apply to any system that involves two passive scalars riding 
on a flow $\bu$, such as $(q,\,\theta)$ for the stratified, rotating Euler and Navier-Stokes 
equations, while on the other hand both the curl of the Lamb vector and the helicity density 
$\lambda = \bom\cdot\bu$ for incompressible flow also possess this behaviour. The significant 
feature is that we are one gradient higher on $\bom$ itself. The embedded gradient of $\bom$ 
within $\bdB$ has been identified as the basis of a test for Euler codes, as explained in 
\S\ref{eultest}. 
\par\smallskip
That the Lamb vector fits into this stretching and folding picture is a surprise.  It is 
generally associated with studies in jet-noise in aero-acoustics and its natural context is 
compressible flows in which wave motion is observed, However, because it is the kinematic 
nonlinearity of fluid flow its evolution is important in also characterizing incompressible 
fluid motion. The stretching and folding process turns out to fit into an electro-magnetic 
analogy. We hope this analogy may become useful in transferring methods of mimetic difference 
schemes that are highly developed for electro-magnetic  applications into the arena of Eulerian 
fluid dynamics. Mimetic methods are reviewed, for example, in Lipnikov, Shashkov \& Yotov (2009).
\par\vspace{3mm}\noindent
\textbf{Acknowledgements:} The authors thank Colin Cotter, Charles Doering, Edriss 
Titi and Claude Bardos for helpful discussions.


\end{document}